\documentclass[twocolumn]{article}
\usepackage{natbib}
\usepackage{graphicx}
\usepackage{amssymb}
\usepackage{widetext}

\begin{document}

\title{Ice loss from the interior of small airless bodies according to an idealized model}
\author{Norbert Sch\"orghofer and Henry H.\ Hsieh \\
Planetary Science Institute, Tucson, AZ and Honolulu, HI, USA}
\maketitle

\begin{abstract}
Ice in main belt asteroids and Near Earth Objects (NEOs) is of scientific and resource exploration interest, but small airless bodies gradually lose their ice to space by outward diffusion. Here, we quantitatively estimate the time it takes a porous airless body to lose all of its interior ice, based on an analytic solution for the interior temperature field of bodies in stable orbits. Without latent heat, the average surface temperature, which is lower than the classic effective temperature, is representative of the body interior and hence an appropriate temperature to evaluate desiccation time scales. 
In a spherically averaged model, an explicit analytic solution is obtained for the depth to ice as a function of time and the time to complete desiccation.
Half of the ice volume is lost after 11\% of this time.
A bilobate structure emerges from the strong latitude dependence of desiccation rates. Cold polar regions can harbor subsurface ice, even when the body center does not. 
Latent heat retards ice loss, and we obtain a succinct expression for the temperature difference between the surface and the ice.
In the outer main belt, nearly all bodies 10~km in size or larger should have been able to retain ice in their interiors over the age of the solar system.
Each of the following factors favors the presence of ice inside NEOs: a semi-major axis in the outer belt or beyond, a mantle of very low thermal inertia, a young age, or a small and stable axis tilt.
\end{abstract}

\section{Introduction}

Water ice in the interior of asteroids and Near Earth Objects (NEOs) is of scientific and resource exploration interest \citep[e.g.][]{lodders99,zealey03}. Airless bodies gradually lose their ice to space by outward diffusion through the porous material, and the time-scale of this desiccation determines whether or not a body that initially contained ice still retains some of it in its interior. To address the question of how much ice a body retains after a given time, and at what depths, for given physical properties and environmental conditions, we obtain analytic solutions for 1) the latitude-dependent temperature field inside a fast-rotating and thermally equilibrated spherical body (section \ref{sec:T2d}), and 2) the time-dependent depth to retreating ice for a spherically symmetric body, including the effect of latent heat (section \ref{sec:i1d}).

A number of numerical models have been developed for icy bodies \citep{delbo15}. For example, \cite{guilbert11} have implemented a full three-dimensional model of ice evolution using spherical harmonics decomposition, and \cite{prialnik92} developed a numerical code for the thermal structure and composition of comet nuclei. \cite{schorghofer08a,schorghofer16a} investigated ice loss from the near-surface with an ensemble of one-dimensional models.  
The approach chosen here differs from these previous works in that it seeks analytic solutions for the interior temperature field and ice retreat rates. 
Analytical one-dimensional thermal models have been previously utilized for comets \citep[e.g.][]{klinger81,kuhrt84,mckay86}.
To obtain these solutions, simplifying assumptions are necessary, but because all parameter dependencies remain transparent, the results provide insight for a wide range of scales and parameters.

\begin{table}[tbh!]
  $D$ ... vapor diffusivity \\
  $L$ ... specific latent heat of water ice sublimation \\
  $\bar Q$ ... mean annual insolation \\
  $T$ ... temperature \\
  $\bar T$ ... temperature averaged over surface area and orbit \\
  $R$ ... body radius \\
  $a$ ... semi-major axis of orbit \\
  $a_\ell$ ... coefficients in a series expansion \\
  $k_B$ ... Boltzmann constant \\
  $r$ ... distance from body center \\
  $r_i$ ... distance of ice table from body center \\
  $t$ ... time \\
  $t_D$ ... time to complete ice loss \\
  $\Gamma$ ... Gamma function \\
  $\zeta$ ... mean free path of water molecules \\
  $\theta$ ... zenith angle (polar coordinate) \\
  $\rho_s$ ... saturation vapor density \\
  $\rho_v$ ... vapor density 
  \caption{Frequently used variables}
\end{table}

\section{Temperature in body interior}
\label{sec:T2d}
\subsection{Problem formulation}

The amplitude of a periodic surface temperature oscillation decays exponentially with depth.
A few such diurnal skin depths below the surface, the temperature varies little throughout a solar day, and below a few seasonal skin depths, it varies little even throughout an orbit around the sun. The spatial domain can be decomposed into a thin spherical shell, where lateral conduction is negligible, and the interior, where the temperature is cylindrically symmetric around the rotation axis of the body.  The temperature in the deep interior of a body can hence be described by a two-dimensional solution, and does not require a three-dimensional description.

If the orbit of the asteroid does not change with time, the temperature in the deep interior is constant as well, and is a function of distance from the body center $r$ and zenith angle or co-latitude $\theta$, defined by the rotation axis of the body. The heat equation for a stationary temperature distribution $T(r,\theta)$ is
\begin{equation}
\nabla^2 T = 0
\label{eq:laplace}
\end{equation}
Throughout this work, it is assumed that the thermal conductivity is spatially uniform.
For small bodies radiogenic heating is negligible beyond the early solar system.

We can estimate how long it takes a body to thermally equilibrate.  In a one-dimensional geometry, the time-scale for the propagation of a heat wave is $R^2/(2\kappa)$, where $\kappa$ is thermal diffusivity. In a spherical geometry, it proceeds faster, so we take $t_T = R^2/(6\kappa)$ as the equilibration time scale, where $R$ is the body radius.
For pure solid ice, the thermal conductivity would be $k\approx 3$~W/m\,K \citep{crc}. Assuming $k=1$~W/m\,K, a density $\rho=2000$~kg/m$^3$, and a heat capacity of $c=1000$~J\,kg\,K for the regolith-ice mixture, the thermal diffusivity is $\kappa = k/(\rho c) \approx 5\times 10^{-7}$~m$^2$/s.
For 2$R$=1 km, $t_T \approx 3\times 10^3$~yrs.  Nearly all icy main belt asteroids should be thermally equilibrated, whereas dynamically young comets will not be thermally equilibrated.

We also numerically estimate the thickness of the shell where temperature changes over one orbit. The thermal skin depth is given by $\delta = \sqrt{k P /(\pi\rho\c)}$, where $P$ is period. For a typical orbital period of $P=5$~yr, a dry layer conductivity of $k=0.1$~W/m\, K, and a silicate heat capacity of $c=500$~J/kg\,K, the skin depth is about 2~m.
The idealized model is valid for bodies with a radius much larger than this seasonal skin depth.

\subsection{Solution ansatz}

In spherical coordinates with cylindrical symmetry, the two-dimensional Laplace equation (\ref{eq:laplace}) becomes
\begin{equation}
{\partial^2 T\over\partial r^2} + 
\frac{2}{r} {\partial T\over\partial r} + {1\over r^2 \sin\theta} {\partial\over\partial\theta} \left(\sin\theta {\partial T\over\partial\theta} \right) = 0
\label{eq:3dlap}
\end{equation}

The partial differential equation (\ref{eq:3dlap}) can be decomposed with Legendre polynomials $P_\ell$. The temperature is expanded into orthogonal trigonometric polynomials
\begin{equation}
T(r,\theta) = \sum_{\ell=0}^\infty T_\ell(r) P_\ell (\cos\theta)
\label{eq:orthoexpansion}
\end{equation} 
The eigenfunction property \citep{jeffrey}
\begin{equation}
{\partial\over\partial\theta} \left(\sin\theta {\partial P_\ell \over\partial\theta} \right) = - \ell(\ell+1) P_\ell
\end{equation}
decouples equation (\ref{eq:3dlap}) into
\begin{equation}
\frac{d^2 T_\ell}{dr^2} 
+ \frac{2}{r}{d T_\ell\over d r} - {\ell(\ell+1)\over r^2} T_\ell  = 0
\label{eq:rode}
\end{equation}
which has solutions 
\begin{equation}
T_\ell(r) = A_\ell \left( {r\over R}\right) ^\ell + B_\ell \left({R\over r}\right)^{\ell+1}
\label{eq:gensolAB}
\end{equation}
The radial dependence is expressed relative to the body radius $R$.
The negative powers of $r$ have to vanish, $B_\ell=0$. To make the coefficients unitless, a constant temperature $\bar T$ can be factored,
\begin{equation}
T_\ell(r) = \bar T a_\ell \left( {r\over R}\right) ^\ell
\label{eq:TofC}
\end{equation}
The temperature at the center of the body is $T(0)=a_0 \bar T$.

From the orthogonality relation
\begin{equation}
\int_{-1}^{1} P_m(x) P_n(x) dx = \frac{2}{2n+1} \delta_{mn}
\end{equation}
and (\ref{eq:orthoexpansion}) we have
\begin{equation}
T_n(r) = \frac{2n+1}{2} 
\int T(r,\theta) P_n(\cos\theta) d\cos \theta 
\label{eq:getcoeffs}
\end{equation}
which provides a way to calculate $a_n=T_n(R)/\bar T$ from the zonally averaged surface temperature $T(R,\theta)$.

For a body in thermal equilibrium, the total radial heat flux through a sphere around the center must vanish, at any depth, 
\begin{equation}
0 = \int \frac{\partial T}{\partial r} dS = \frac{d}{dr} \int T dS
\end{equation}
where the integral is over the surface of the sphere. Hence the temperature averaged over the surface of the body, $\bar T$, must be the same as the temperature at the center of the body. 
With this relation, the solution can be written as
\begin{equation}
T(r,\theta) = \bar T \sum_{\ell=0}^\infty a_\ell \left( {r\over R}\right) ^\ell P_\ell(\cos\theta)
\label{eq:fullsolution}
\end{equation}
where $a_\ell = T_\ell(R)/\bar T$ and $a_0=1$.

If the temperature is symmetric on the two hemispheres, the odd coefficients vanish, and 
\begin{equation}
T(r,\theta) = \bar T  
\left[ 1 + a_2 \left({r\over R}\right)^2 \left( \cos^2\theta-{1 \over 2}\right) + ... \right]
\end{equation}
The radial dependence involves only $r^0, r^2, r^4, ...$, which implies that the $\theta$-dependent terms ($\ell>0$) diminish rapidly for small $(r/R)$.
The slowest non-constant term decays as $r^2$. Hence, the depth-scale of this temperature change is determined by $r/R \approx 1/\sqrt{2}$, i.e.\ roughly 3/10th body radii below the surface.

\subsection{Fast rotator model as surface boundary condition}
A body's effective temperature is defined by
\begin{equation}
\epsilon\sigma T_{\rm eff}^4 = \frac{1}{4} (1-A) \bar Q 
\label{eq:Teff}
\end{equation}
where $\sigma$ is the Stefan-Boltzmann constant, $A$ albedo, and $\bar Q$ the annual mean insolation (for a circular orbit the solar constant at the appropriate distance from the sun). The factor of 4 represents the ratio of the surface of a sphere to its cross-sectional area.

The fast rotator model \citep{lebofsky89} assumes the rotation of the body is sufficiently fast so that the temperature does not change with geographic longitude (local time), or, equivalently that the thermal inertia is infinite. This often is not a good approximation, but it is a well-known end-member thermal model of asteroids.  It applies best to rocky bodies (high thermal inertia) that spin fast.

In the following it is also assumed that the body has zero axis tilt, so the incoming flux is simple to calculate. 
The solar flux which falls onto each infinitesimal latitude band at co-latitude $\theta$ is distributed over a circle of latitude,
\begin{equation}
\epsilon\sigma T^4 = \frac{1}{\pi} (1-A) \bar Q \sin\theta 
\end{equation}
The factor of $\pi$ represents the ratio of the circumference of a circle to its diameter.
In this simple situation, the surface temperature is
\begin{equation}
T(R,\theta)=T_{eq} \sin^{1/4} \theta
\end{equation}
and the temperature at the equator is
\begin{equation}
T_{eq} = \left[ \frac{(1-A) \bar Q }{\pi \epsilon\sigma} \right]^{1/4}
= \frac{\sqrt 2}{\pi^{1/4}}  T_{\rm eff} \approx 1.062 \, T_{\rm eff}
\end{equation}

The temperature averaged over the surface area is
\begin{eqnarray}
\bar T &=& \frac{1}{2} \int^{\pi}_{0} T \sin\theta d\theta
= \frac{T_{eq}}{2}  \int^{\pi}_{0} \sin^{5/4}\theta d\theta \\
&=& \frac{\sqrt\pi}{10} {\Gamma(1/8)\over\Gamma(5/8)} T_{eq} 
\approx 0.93087 \, T_{eq} 
\label{eq:Tbar}
\end{eqnarray}
where $\Gamma$ denotes the Gamma function.
Or, expressed in terms of $T_{\rm eff}$,
\begin{equation}
\bar T = \frac{\pi^{1/4}}{5\sqrt{2}} {\Gamma(1/8)\over\Gamma(5/8)} T_{\rm eff}
\approx 0.98882 \, T_{\rm eff}
\label{eq:TfromTeff}
\end{equation}
The average surface temperature (area-weighted) is about 1\% lower than the effective temperature.

If the only source of energy is the sun, then it is always the case that $T_{\rm eff} \geq \bar T$. The effective temperature corresponds to the energy balance for a uniform surface temperature, and because the outward infrared radiation goes as the fourth power of temperature (Stefan-Boltzmann law), any deviation from uniformity leads to additional energy loss. For dust covered bodies (low thermal conductivity), this temperature can easily be 20~K lower \citep{schorghofer08a,schorghofer16a}, which has an enormous impact on sublimation rates.

\subsection{Solution for interior temperature}
We now determine the coefficients $a_\ell$ from the surface temperature.
Based on eqs.\ (\ref{eq:TofC},\ref{eq:getcoeffs}),
\begin{eqnarray}
a_\ell \bar T &=& \frac{2\ell +1}{2} T_e I_\ell
\label{eq:ctoT}\\
I_\ell &=& \int_{-1}^1 (1-x^2)^{1/8} P_\ell(x) dx
\label{eq:Inu}
\end{eqnarray}
where $x=\cos\theta$.
Odd coefficients vanish, due to hemispheric symmetry.

\begin{widetext}
A special case of an integration formula given in \cite{jeffrey}, 7.132 is
\begin{equation}
\int_{-1}^1 (1-x^2)^{\lambda-1} P_\nu(x) dx =
{ \pi \Gamma(\lambda) \Gamma(\lambda) \over
\Gamma\left(\lambda + \frac{\nu}{2}+\frac{1}{2} \right) \Gamma\left(\lambda-\frac{\nu}{2}\right) \Gamma\left(\frac{\nu}{2} + 1 \right) \Gamma\left( -\frac{\nu}{2} + \frac{1}{2}\right) }
\end{equation}
For $\lambda=9/8$, as needed in our case (\ref{eq:Inu}),
\begin{eqnarray}
I_\nu = \int_{-1}^1 (1-x^2)^{1/8} P_\nu(x) dx &=&
{ \pi \Gamma^2\left(\frac{9}{8} \right) \over
\Gamma\left(\frac{13}{8} + \frac{\nu}{2} \right) \Gamma\left(\frac{9}{8}-\frac{\nu}{2}\right) \Gamma\left(\frac{\nu}{2} + 1 \right) \Gamma\left( -\frac{\nu}{2} + \frac{1}{2}\right) } 
\\ &=& {\pi \Gamma^2\left(\frac{1}{8} \right) \over
(5 + 4\nu) (1-4\nu) \Gamma\left(\frac{5}{8} + \frac{\nu}{2} \right) \Gamma\left(\frac{1}{8}-\frac{\nu}{2}\right) 
\Gamma\left(\frac{\nu}{2} + 1 \right)  \Gamma\left(-\frac{\nu}{2} + \frac{1}{2}\right) }
\label{eq:bigint2}
\end{eqnarray}
where we have used $\Gamma(x+1) = x \Gamma(x)$.
\end{widetext}

For $\nu=0$,
\begin{equation}
I_0 = {\sqrt\pi \over 5} { \Gamma\left(\frac{1}{8} \right) \over \Gamma\left(\frac{5}{8}  \right) }
\end{equation}
From comparison with (\ref{eq:Tbar},\ref{eq:ctoT}), it is apparent that $a_0=1$, as we already determined above.
For $\nu=2$, 
\begin{equation}
I_2 = - {\sqrt\pi \over 130 } {\Gamma\left(\frac{1}{8}\right) \over \Gamma\left(\frac{5}{8}\right)  }
\end{equation}
The recursion relation, obtained from (\ref{eq:bigint2}),
\begin{equation}
I_{\nu+2} = { (4\nu - 1) (\nu+1)  \over (13 + 4\nu) (\nu + 2) } I_\nu
\label{eq:Irecursion}
\end{equation}
involves no special functions. The first is
\begin{equation}
I_2 = - {  1  \over 26 } I_0
\end{equation}
Only the first recursion coefficient is negative, which means $I_0>0$, but $I_{2n}<0$ for $n>0$.
Combined with (\ref{eq:ctoT}), the recursion for the coefficients $a_n$ is
\begin{eqnarray}
a_n &=& \frac{2n+1}{2n-3} \frac{4n-9}{4n+5} \frac{n-1}{n} a_{n-2}
\label{eq:Crecursion} 
\end{eqnarray}
Numerically,
$a_0 = 1$, $a_2 = -5/26 \approx -0.1923$, $a_4 = - 9/104 \approx -0.0865$, $a_6 \approx -0.0539$, ....

\begin{figure}[tbh!]
a)\\
\includegraphics[width=7.5cm]{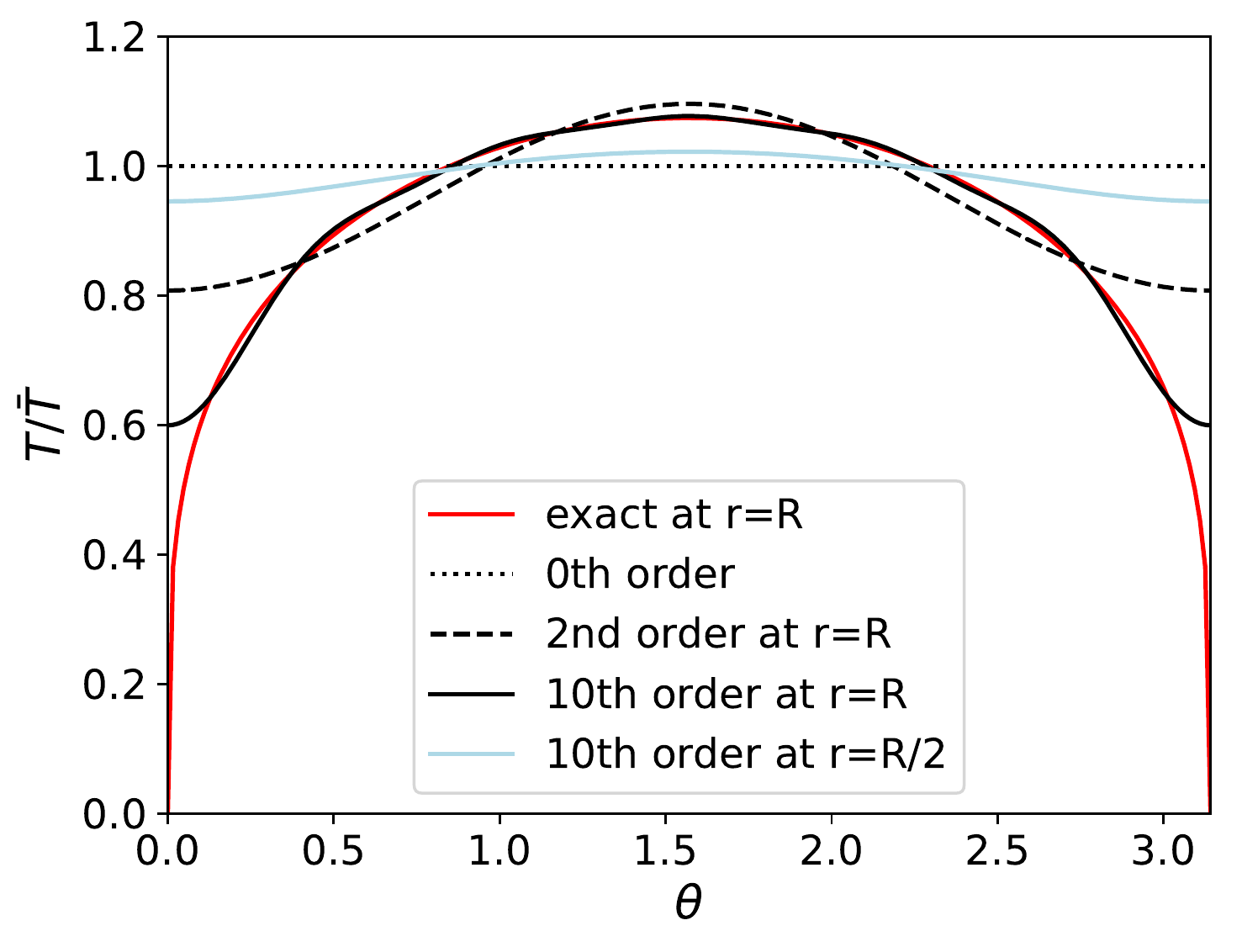}\\
b)\\
\includegraphics[width=7.5cm]{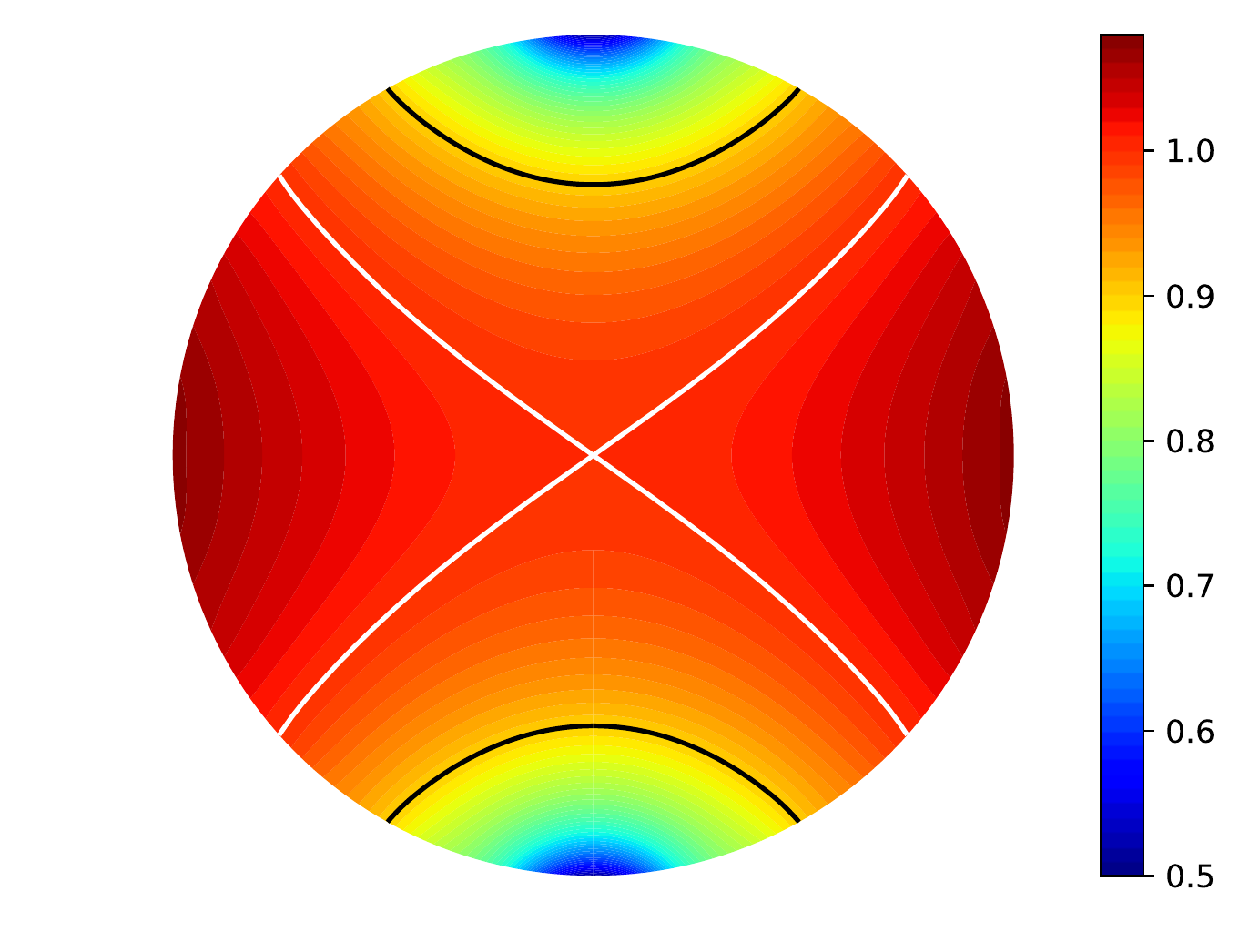}
\caption{
Temperature distribution of a thermally equilibrated spherical fast-rotator with zero axis tilt.
a) Approximation of surface temperature $T\propto (\sin\theta)^{1/4}$ with a series of Legendre polynomials. At half body radius ($r=R/2$), the $\theta$-dependence is small.
b) Temperature distribution in the body interior, based on 20th order analytic solution. 
Shown is the cross-section $T(r,\theta)$, and the solution is cylindrically symmetric around the rotation axis.
The color scale shows $T/\bar T$, where $\bar T$ is the temperature of the object averaged over its entire surface area and orbit. The white contours show $T/\bar T=1$, and the black contours $T/\bar T=0.9$.
\label{fig:Tsolution}}
\end{figure}

Figure~\ref{fig:Tsolution}a shows the surface temperature for increasing orders in the series expansion, $T(R,\theta)/\bar T=\sum_{\ell=0}^n a_\ell P_\ell(\theta)$, compared to the exact solution, $T=T_e \sin^{1/4}\theta$. The deviation is largest at the poles, and that is welcome since the zero polar temperatures are an unrealistic aspect of the idealized model. The solution at $T(R/2,\theta)$ illustrates how quickly the temperature homogenizes with depth.

Figure~\ref{fig:Tsolution}b shows the interior temperature distribution as a function of $r/R$ and $\theta$. The cold polar areas are shallow, consistent with the above estimate that lateral temperature changes decay with a depth-scale of 3/10th of the body radius.

\subsection{Numerical values for interior temperatures}
Since the surface area averaged temperature $\bar T$ determines the interior temperature, we discuss here how $\bar T$ is related to orbital geometry and physical properties of the body.

\begin{figure}[tbh!]
a)\\
\includegraphics[width=7.5cm]{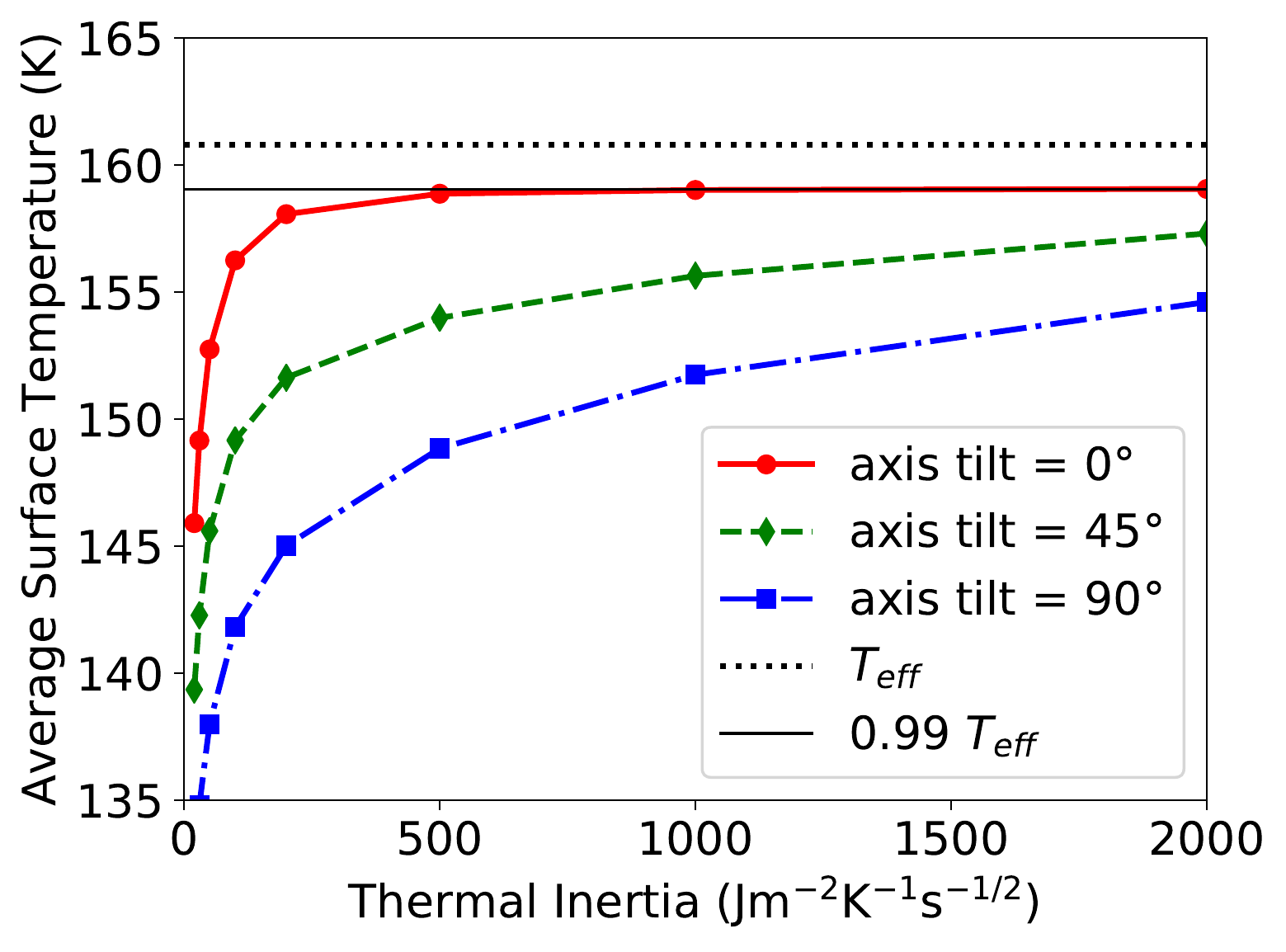}\\
b)\\
\includegraphics[width=7.5cm]{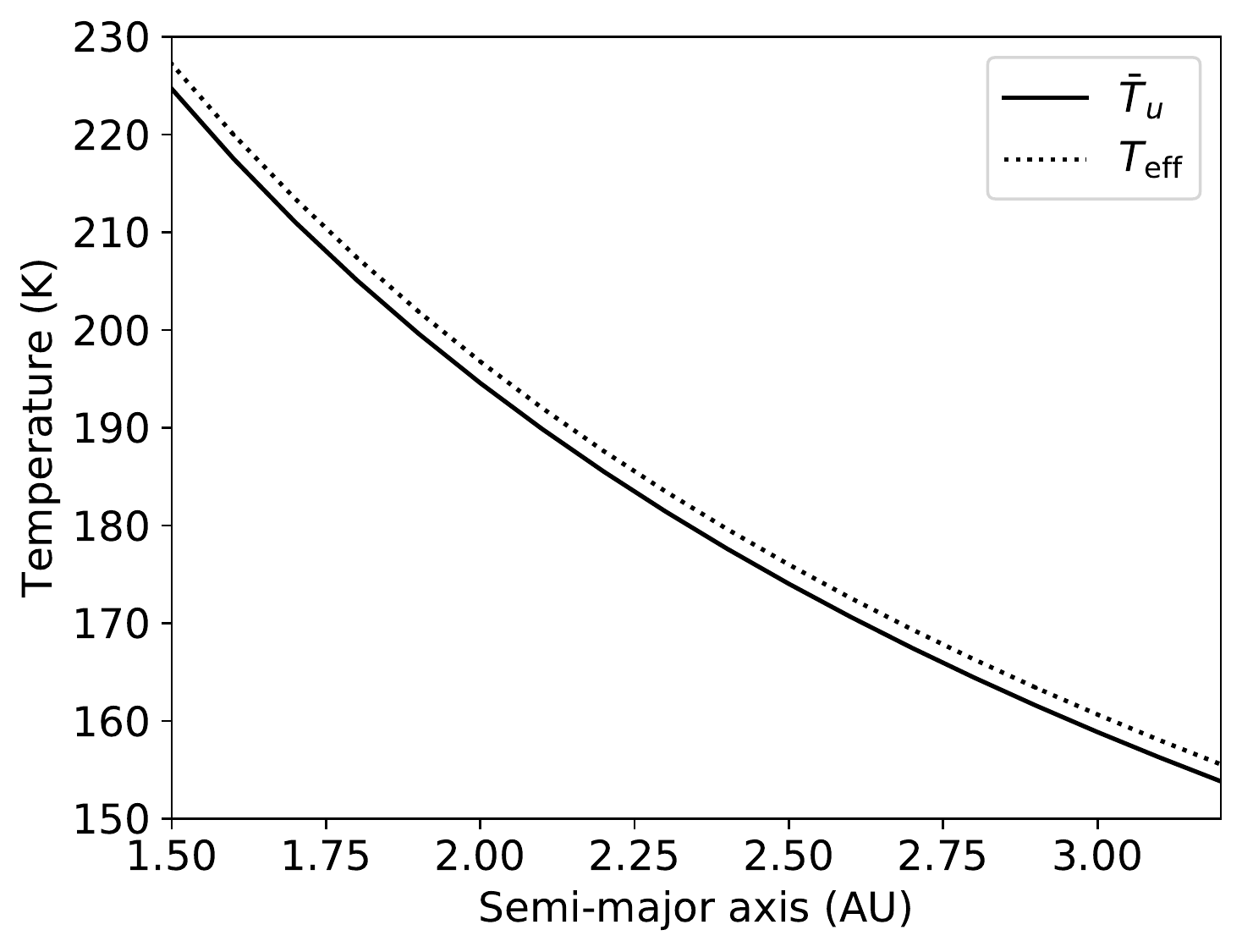}
\caption{
a) Area-averaged surface temperature $\bar T$ according to numerical thermal model calculations for $a=3$~AU and a rotation period of 6~hr. A thermal inertia of 2000~J~m$^{-2}$~K$^{-1}$~s$^{-1/2}$ would represent a single block of rock or ice. Small thermal inertia is representative of dust-sized particles.
b) Estimated maximum interior temperature $\bar T_u$. The actual interior temperature (surface temperature) can be lower than $\bar T_u$ because of axis tilt or low thermal inertia.
For both panels, an albedo of 5\%, an emissivity of 0.95, and a circular orbit were assumed.
\label{fig:Tfroma}}
\end{figure}

Figure \ref{fig:Tfroma}a shows the results of numerical surface temperature calculations, carried out with a numerical 1D model for a range of latitudes that cover the entire globe. This standard thermophysical model is described in \cite{github}. The figure shows $\bar T$ as a function of thermal inertia. For zero axis tilt and high thermal inertia this temperature must agree with eqs.\ (\ref{eq:Teff},\ref{eq:TfromTeff}).  At low thermal inertia, average temperature decreases significantly \citep{schorghofer16a}. A higher axis tilt also leads to lower average surface temperature, because the seasonal temperature amplitude becomes larger, and a larger amplitude leads to more thermal radiation to space.

The influence of orbital eccentricity $e$ can to some extent be assessed analytically. The mean annual insolation $\bar Q$ is proportional to $1/\sqrt{1-e^2}$ \citep[e.g.][]{ward74a,klinger81}.
Hence,
\begin{equation}
\bar Q = {1\over \sqrt{1-e^2}} \frac{S_o}{a^2}
\end{equation}
where $S_o$ is the solar constant at 1~AU and $a$ the semi-major axis.
The equivalent circular orbit would have radius $a (1-e^2)^{1/4}$. 
Even for moderate eccentricity, this is not a large change compared to a circular orbit.
This does not imply that $\bar T$ has the same dependence on $e$ \citep{rubincam04}, but it roughly captures the magnitude of the effect.

For a fast rotator with zero axis tilt $\bar T = 0.99 T_{\rm eff}$, where the effective temperature (\ref{eq:Teff}) is calculated from the annual mean insolation. We call this temperature $\bar T_u$ because it appears to be an upper bound on $\bar T$,
\begin{equation}
\bar T_u = 0.9888 \left[{1\over \sqrt{1-e^2}} \frac{1-A}{4\epsilon\sigma} \frac{S_o}{a^2} \right]^{1/4}
\end{equation}
$\bar T$ may be colder due to axis tilt or low thermal inertia (Fig.~\ref{fig:Tfroma}), but based on the numerical evidence presented in Fig.~\ref{fig:Tfroma}a, $\bar T_u \geq \bar T$. The desiccation timescale obtained from $\bar T_u$ will correspondingly be a conservative estimate.

\section{Model of ice retreat}
\label{sec:i1d}
Here we consider long-term loss of ice by vapor diffusion through the overlying porous layer.

\subsection{Equation for ice retreat}

First we consider a fully spherically symmetric situation to calculate the mass loss. Since vapor moves much faster than ice and heat, it also equilibrates faster, so the vapor flux is stationary.
At the ice interface ($r=r_i$) the vapor flux is given by
\begin{equation}
J(r_i) = \rho_i \frac{dr_i}{dt} 
\label{eq:retreatdef}
\end{equation}
where $\rho_i$ is the density of ice and $r_i$ is the radial coordinate of the receding ice table.
The vapor flux must be proportional to $1/r^2$,
\begin{equation}
J(r) = -D \frac{d\rho_v(r)}{d r} = \left( \frac{r_i}{r} \right)^2 J(r_i)
\end{equation}
where $D$ is the vapor diffusivity and $\rho_v$ the vapor density.
Upon integration,
\begin{equation}
\rho_v(r) = {r_i^2} \frac{J(r_i)}{D} \left(\frac{1}{r} - \frac{1}{R} \right)
\label{eq:misc}
\end{equation}
The boundary values for the vapor density are $\rho_v(r=R)=0$ on the body's surface and $\rho_v(r=r_i)=\rho_s$ at the ice interface, where $\rho_s$ is the saturation vapor density at temperature $T(r_i)$. 
Hence
\begin{equation}
\rho_v(r) = \rho_s { \frac{1}{r} - \frac{1}{R} \over \frac{1}{r_i} - \frac{1}{R} }
\end{equation}
Taking the derivative thereof at $r=r_i$,
\begin{equation}
J(r_i) = \frac{D \rho_s}{ r_i (1- r_i/R)} 
\label{eq:flux3}
\end{equation}
(If $r_i$ was close to $R$, $r_i=R-z$, this would reduce to $J(r_i) = D \rho_s / z $, as it should.)
Combining (\ref{eq:flux3}) with (\ref{eq:retreatdef}),
\begin{equation}
\frac{dr_i}{dt} =  D \frac{\rho_s}{\rho_i} \frac{1}{ r_i (1- r_i/R)}
\label{eq:drdt}
\end{equation}
$\rho_s$ only depends on temperature, so this is a differential equation that can be integrated over time.

\subsection{Solution for spherically averaged model}
For constant temperature (1-dimensional spherical model), $\rho_s(T)$ becomes a constant, $\rho_s(\bar T)$, and the differential equation (\ref{eq:drdt}) integrates to
\begin{equation}
  r_i \left(1- \frac{r_i}{R} \right) dr_i =  D \frac{\rho_s}{\rho_i} dt
\end{equation}
and upon integration from $R$ to $r_i$, and 0 to $t$,
\begin{equation}
\frac{R^2}{6} - \frac{r_i^2}{2} + \frac{r_i^3}{3R} = D \frac{\rho_s}{\rho_i} t
\label{eq:tofr}
\end{equation}
The time to complete desiccation $t_D$ is given by $r_i=0$,
\begin{equation}
t_D = \frac{R^2}{6D} \frac{\rho_i}{\rho_s(\bar T)}
\label{eq:tD}
\end{equation}
In a planar situation the factor of 6 would be 2 instead.

In non-dimensional variables $r'=r/R$ and $t'=t/t_D$, (\ref{eq:tofr}) becomes
\begin{equation}
1 -  3  r_i'^2 + 2 r_i'^3 = t'
\label{eq:tofrprime}
\end{equation}
The ice has receded to half the body radius ($r_i'=1/2$) exactly at $t=t_D/2$. Only 1/8th of the ice volume is left at this stage. Half the ice volume is lost at $t/t_D = 2 -3/\sqrt[3]{4}= 0.11$, that is, after 11\% of the total desiccation time. Two thirds are lost after $t/t_D=5/3-3^{1/3}$ or 22\% of the time.

\begin{figure}
\includegraphics[width=7.5cm]{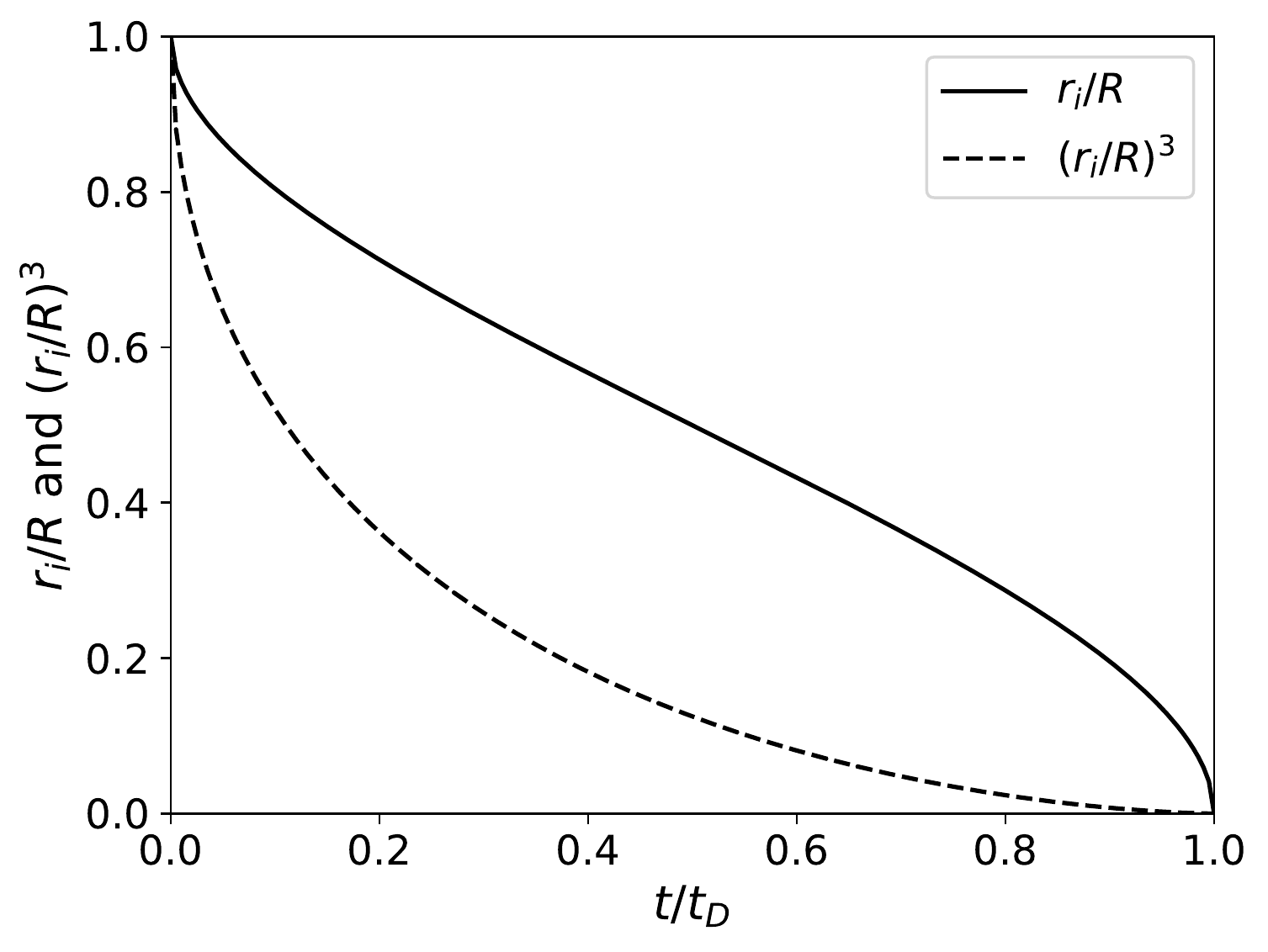}
\caption{Time dependence of ice retreat in a spherically averaged model according to equation (\ref{eq:roft}). The retreat is also shown in terms of relative volume retained.}
\label{fig:cubic}
\end{figure}

The inverse of (\ref{eq:tofrprime}), $r'_i(t')$, is the solution to a cubic equation.
With $r_i'= u + 1/2$ it turns into the form
\begin{equation}
u^3 - \frac{3}{4} u + \frac{1-2t'}{4} =0
\end{equation}
The discriminant is obtained as
\begin{equation}
\Delta 
= \frac{27}{4} t' (1-t') \geq 0 
\end{equation}
and hence there are three real roots that can be expressed by trigonometric functions.
The roots are
\begin{equation}
u_k = \cos\left(\frac{1}{3} \arccos(2t'-1) - \frac{2\pi k}{3} \right) \quad k=0,1,2
\end{equation}
The physically relevant of the three solutions turns out to be $k=1$.
Undoing the reductions,
\begin{equation}
\frac{r_i(t)}{R} = \frac{1}{2} + \cos\left(\frac{1}{3} \arccos\left(2\frac{t}{t_D}-1\right) - \frac{2\pi}{3} \right)
\label{eq:roft}
\end{equation}
Figure~\ref{fig:cubic} shows this universal solution. The ice recedes fastest at the beginning and at the end. The retained ice volume is proportional to $(r_i/R)^3$, and this fraction is also plotted in Figure~\ref{fig:cubic}.

The depth to the ice table is
\begin{equation}
R - r_i(t) = \frac{R}{2} - R\cos\left(\frac{1}{3}\arccos\left(2\frac{t}{t_D} - 1 \right) - \frac{2\pi}{3} \right)
\label{eq:doft}
\end{equation}
For small $t$ this must reduce to the planar case.
Indeed, the series expansion is
\begin{equation}
R - r_i = R \sqrt{ t \over 3 t_D} + O(t) = 
\sqrt{2 D t \frac{\rho_s}{\rho_i}} + O(t)
\end{equation}

The retreat rate as a function of time is
\begin{equation}
  \frac{d r_i}{dt} = \frac{R}{3} {\cos\left(\frac 1 3 \arcsin\left( 1 - 2 \frac{t}{t_D} \right) \right) \over \sqrt{ t(t_D -t)} }
\end{equation}
The outgassing rate is
$- \rho_i 4\pi r_i^2 (d r_i / dt)$. It is largest at the beginning and decreases monotonically toward zero.  For example, at $t = t_D/2$ the outgassing rate is $\rho_i (2\pi /3) R^3/t_D$.

\subsection{Desiccation time}
To determine absolute desiccation time, we need to choose values for body size $R$, body temperature $\bar T$, and vapor diffusivity $D$. The saturation vapor density $\rho_s(\bar T)$ can be calculated from temperature with established formulae. 
The saturation vapor pressure is \citep{bryson74,sack93,murphy05}
\begin{equation}
\ln p_s =  b {\,-} \frac{m L}{k_B T} 
\label{eq:psv}
\end{equation}
Here, $m$ is the mass of a water molecule, $k_B$ the Boltzmann constant, and $L$ the specific latent heat of ice.
For water ice, $m L/k_B \approx 6140$~K (0.53~eV) and $b\approx 28.9$ \citep{murphy05}.

The vapor diffusion coefficient $D$ depends on pore sizes (related to grain size), pore structure, and temperature. In a free ideal gas it would be $(1/3) \bar v_{\rm th} \zeta$, where $\bar v_{\rm th}$ is the mean thermal speed and $\zeta$ the mean free path. For a Maxwell distribution of velocities $\bar v_{\rm th} = \sqrt{8\pi k_B T/\pi m}$. In the current context, the water molecules migrate by adsorption and desorption rather than collisions between water molecules.
In this case, a Maxwell distribution may no longer be applicable, but the change to the
mean thermal speed is negligible compared to other uncertainties.
The pore size $\zeta$ in the interior is related to grain size, often a major unknown \citep{asphaug02rev}.
\cite{herique18} indicate regolith on asteroidal surfaces has a size range from microns to a meter, and the internal aggregate may be coarser than that.
We choose a value of $\zeta=1$~cm and note that $t_D$ is proportional to $\zeta$ and $R$ is proportional to $\sqrt\zeta$.
In a porous medium, $D$ also depends on porosity $\phi$, but the ice density $\rho_i$ is also proportional to porosity. So to leading order the two porosity factors cancel and we take $\rho_i/\phi = 927$ kg/m$^3$ and $D/\phi =(1/3) \bar v_{\rm th} \zeta$, because we are essentially only choosing a value for $\rho_i/D$.

\begin{figure}[tbh!]
a)\\
\includegraphics[width=7.5cm]{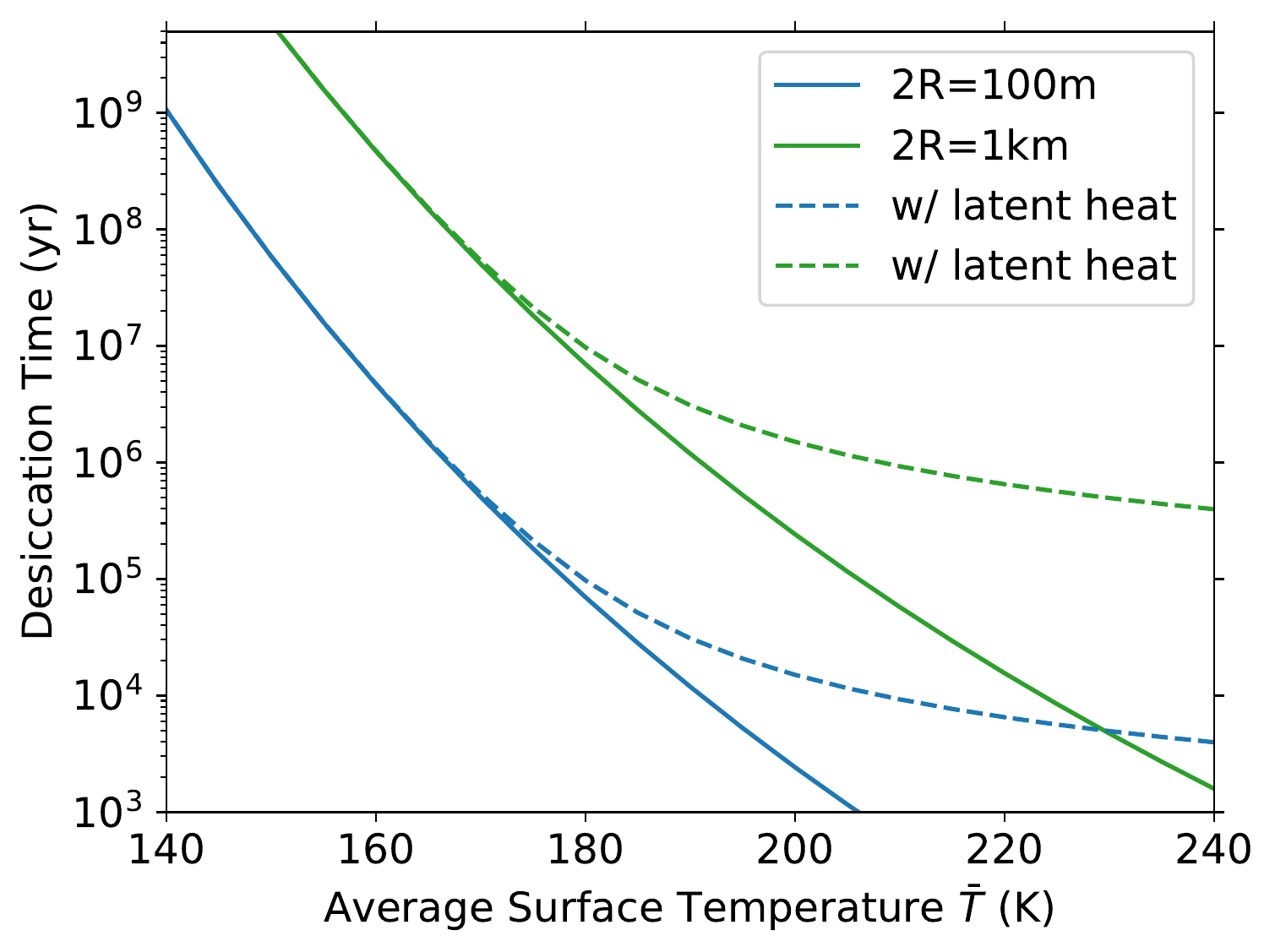}\\
b)\\
\includegraphics[width=7.5cm]{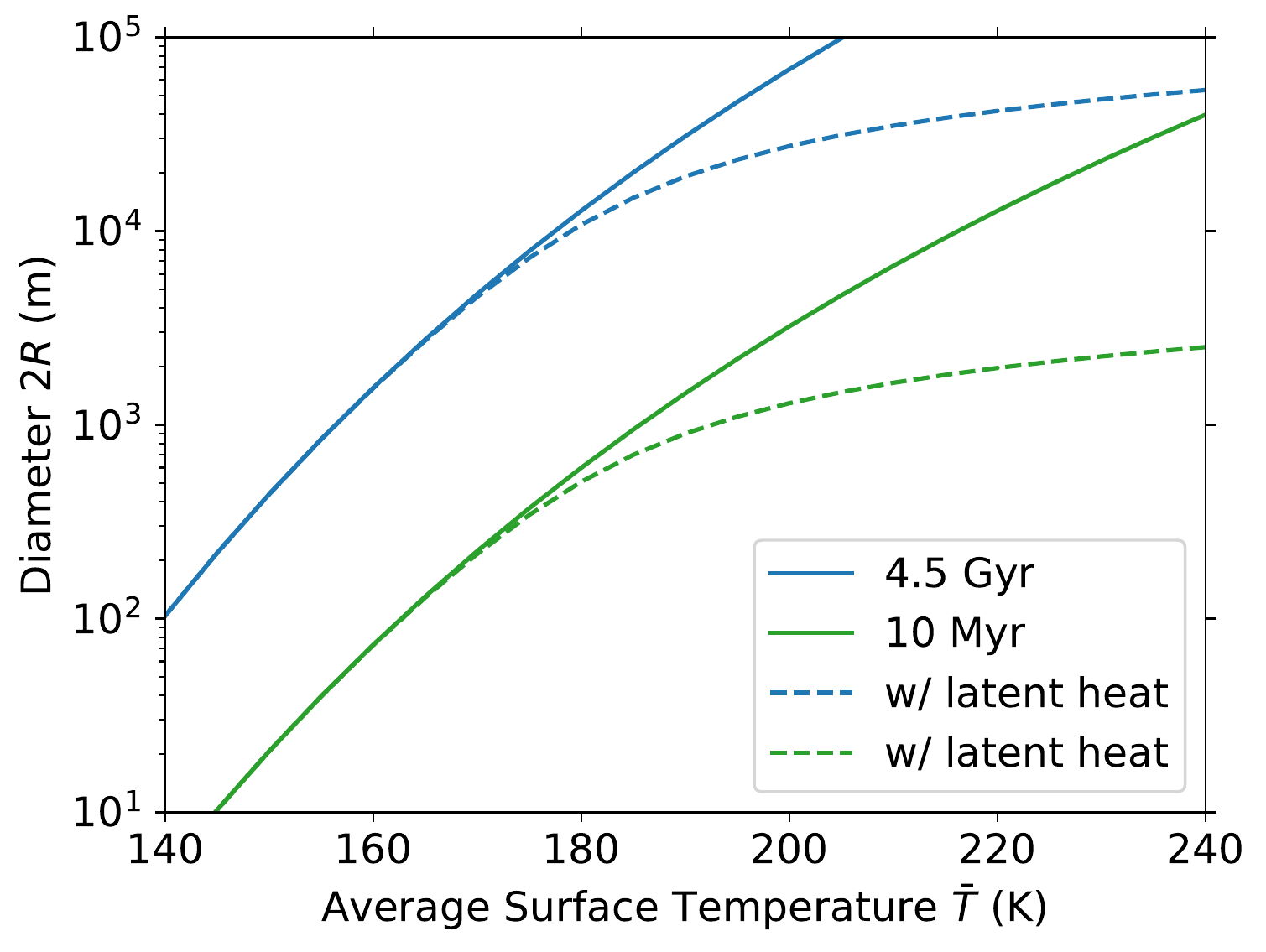}
\caption{a) Desiccation time $t_D$ in Earth years as a function of mean surface temperature $\bar T$ for pore size $\zeta=1$~cm and body diameters $2R$, according to equation (\ref{eq:tD}) (solid lines).
The dashed lines are desiccation times $t_L$ with the latent heat effect included and thermal conductivity of the dry material of $k=0.1$~W/mK, according to equations (\ref{eq:DeltaTfinal}) and (\ref{eq:tL}).
b) Same as (a) but plotting threshold diameter for fixed desiccation times as a function of $\bar T$.
\label{fig:thenumbers}}
\end{figure}

We choose an average surface temperature $\bar T$, a body size $R$, and a vapor diffusion mean free path $\zeta$, and from these calculate vapor diffusivity $D$ and the desiccation time $t_D$. Figure~\ref{fig:thenumbers}a shows desiccation times as a function of $\bar T$ and for two body diameters.  

Figure~\ref{fig:thenumbers}b is based on the same equations as Figure~\ref{fig:thenumbers}a, but instead of desiccation time for fixed body diameters, it shows the threshold body diameter for fixed desiccation times (ages). Bodies larger than this threshold diameter have retained some ice in their interior, if they started out as ice-rich objects, while bodies smaller than the threshold diameter can be expected to have lost all ice.
(Here, $\bar T$ can be obtained by any independent means, and this result is not limited to the fast rotator with zero axis tilt that was used for the two-dimensional temperature field.)

\subsection{Latent heat effect}

The latent heat consumed at the ice interface is
\begin{equation}
Q = \rho_i \frac{dr_i}{dt} L
\label{eq:Q}
\end{equation}
where $L$ is the specific latent heat.

We consider a stationary temperature field where the latent heat of sublimation is compensated by heat flux through the ice-free shell. Heat can also be drawn toward the ice table from the sensible heat of the body interior. The following arguments justify that neglecting sensible heat should be a good assumption:
1) By the end of the desiccation process, the sensible heat will have to be returned, so sensible heat acts in part only to redistribute the energy over time.
2) As is well known, the latent heat of melting corresponds to a difference in heat capacity by 70$^\circ$C, and the latent heat of sublimation is about 7 times larger than that of melting, so the sensible heat stored in the entire body is small compared to the latent heat required to sublimate all of the ice.

The stationary solution is, based on (\ref{eq:gensolAB}),
\begin{equation}
T(r) = \left\{
\begin{array}{ll}
A + B/r  \quad\mbox{for}\quad r>r_i \\
C  \quad\mbox{for}\quad r<r_i
\end{array}
\right.
\end{equation}
From the boundary values, it quickly follows that
\begin{equation}
B = - {\bar T-T(r_i) \over \frac{1}{r_i} - \frac{1}{R} }
\end{equation}
The heat flux in the ice-free domain is
\begin{equation}
H = -k \frac{\partial T}{\partial r} = - k \frac{B}{r^2} 
\end{equation}
where $k$ is the thermal conductivity of the ice-free domain.
Balancing the heat flux with the latent heat,
$H(r_i) = Q$,
\begin{equation}
B  = - \frac{Q}{k} r_i^2
\end{equation}
The temperature at the ice table determines the vapor pressure.
The reduction in temperature due to latent heat is
\begin{equation}
\bar T - T(r_i) = \frac{Q}{k} r_i \left(1-\frac{r_i}{R}\right)
\label{eq:DeltaT}
\end{equation}

Combining (\ref{eq:drdt}), (\ref{eq:Q}), and (\ref{eq:DeltaT}) yields a simple relation
\begin{equation}
\Delta T = \bar T - T(r_i) 
= \frac{D}{k} \rho_s L 
\label{eq:DeltaTfinal}
\end{equation}
The temperature difference depends on neither $R$ nor directly on $r_i$, so the ice table is characterized by a single temperature $T_i \equiv T(r_i)$.

The simplicity of (\ref{eq:DeltaTfinal}) can be understood in the following way.
At the ice table, the heat flux balances evaporative cooling:
\begin{equation}
-k \frac{\partial T}{\partial r} = - D \frac{\partial \rho_v}{\partial r} L
\end{equation}
Since temperature $T$ and vapor density $\rho_v$ both obey the Laplace equation in the same geometry, the solutions are geometrically similar (the heat flux $H$ and the vapor flux $J$ are both proportional to $1/r^2$) and the ratio of the local derivatives can be replaced with the ratio of the differences between the respective boundary values, $\Delta T$ and $\Delta\rho_v=\rho_s$,
\begin{equation}
-k {\Delta T} = - D \rho_s L
\end{equation}
This reproduces (\ref{eq:DeltaTfinal}).

With $\rho_s(T(r_i))$, (\ref{eq:DeltaTfinal}) is a nonlinear equation for $T(r_i)$. 
For small differences, the relation can be linearized
\begin{equation}
\Delta\rho_s = \frac{d\rho_s}{dT} \Delta T
\label{eq:Deltarhos}
\end{equation}
The saturation vapor pressure (\ref{eq:psv}) combined with the ideal gas law, 
\begin{equation}
k_B T \rho_s = m p_s
\label{eq:idealgas}
\end{equation}
yields
\begin{equation}
\frac{d\rho_s}{dT} = \frac{\rho_s}{T} \left(\frac{m L}{k_B T} - 1 \right)
\approx \frac{\rho_s}{T} \frac{m L}{k_B T} 
\label{eq:drhosdT}
\end{equation}
Combining (\ref{eq:DeltaTfinal}), (\ref{eq:Deltarhos}), and (\ref{eq:drhosdT}),
\begin{equation}
\frac{\Delta\rho_s}{\rho_s}  \approx
\frac{D}{k} \rho_s \frac{m L^2}{k_B T^2} 
= \frac{D}{kT} p_s \left(\frac{m L}{k_B T} \right)^2
\label{eq:lcriterion}
\end{equation}

Latent heat becomes noticeable when $\Delta\rho_s/\rho_s \approx 1/2$. For $D/\phi=3$~m$^2$/s and $k=0.1$~Wm$^{-1}$K$^{-1}$, the cross-over occurs at $T\approx 180$~K.  In bodies warmer than that, latent heat will slow down the ice loss. Grain size changes both $D$ and $k$ in the same direction, so the ratio $D/k$ is less dependent on grain size than $D$ and $k$ are individually.

Since $\Delta T$ is constant in time, the only change in (\ref{eq:drdt}) is that $\rho_s$ has to be evaluated at $T_i$ instead of $\bar T$, and (\ref{eq:tD}) and (\ref{eq:roft}) are still valid. We rename $t_D$ to $t_L$ when latent heat is taken into account,
\begin{equation}
t_L = \frac{R^2}{6D} \frac{\rho_i}{\rho_s(T_i)}
\label{eq:tL}
\end{equation}

\begin{figure}
\includegraphics[width=7.5cm]{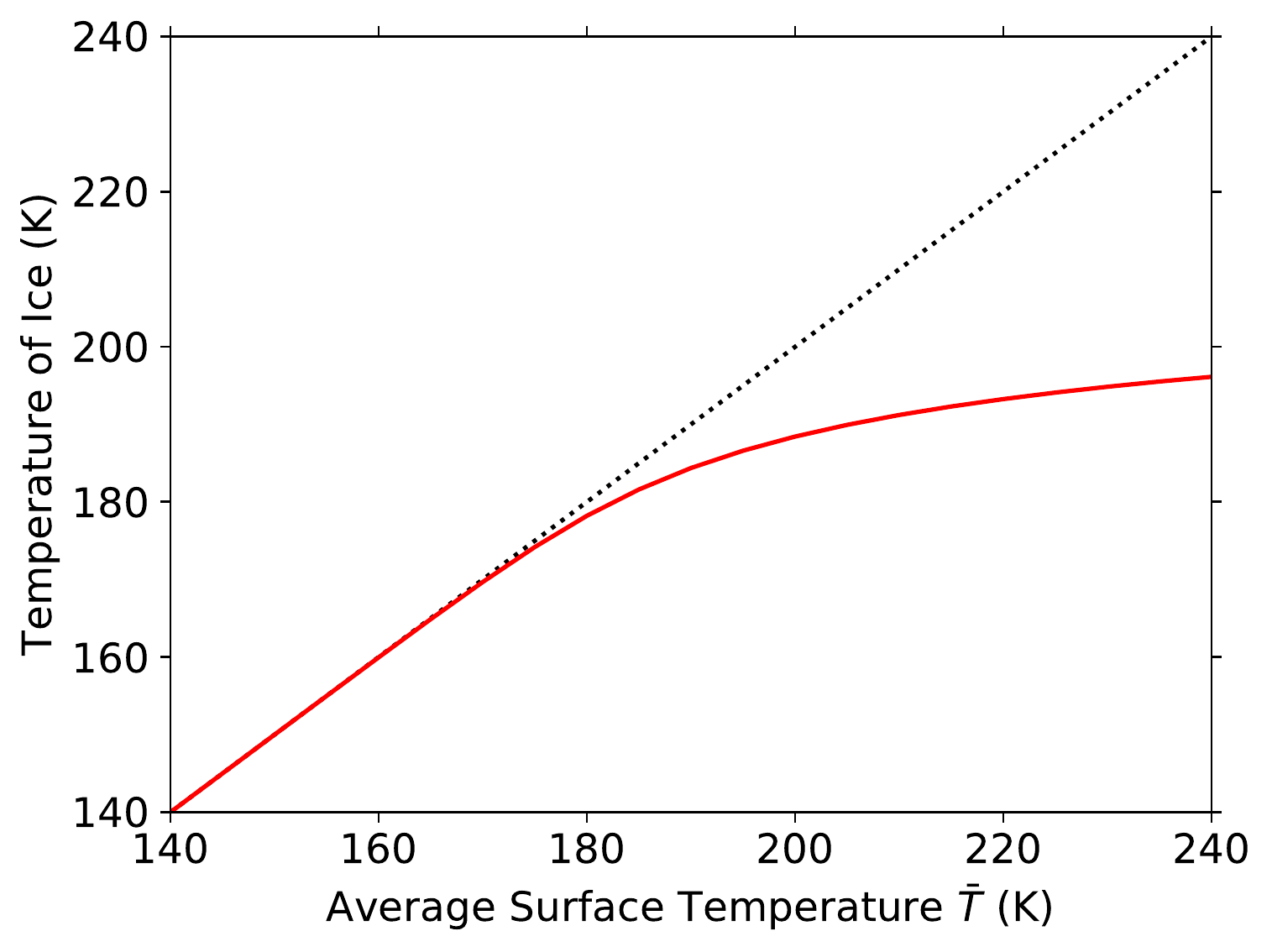}
\caption{The temperature of the ice (red line) versus the average surface temperature $\bar T$ shows the temperature difference between the surface and the ice due to latent heat according to equation (\ref{eq:DeltaTfinal}). The dotted line is a 1:1 relation (no difference). Here, $\zeta=1$~cm and $k= 0.1$~W/m\,K. 
\label{fig:latentTdiff}}
\end{figure}

To obtain $T_i$ the nonlinear equation (\ref{eq:DeltaTfinal}) with (\ref{eq:psv},\ref{eq:idealgas}) has been solved numerically.
Figure~\ref{fig:latentTdiff} shows the temperature difference due to the latent heat effect.
At low temperature the latent heat has no effect.  
As anticipated, for temperatures above about 180~K, latent heat becomes noticeable, and above 190~K it is significant.
At higher temperatures, the ice is significantly cooler than the average surface, which prolongs desiccation times.
Figure~\ref{fig:thenumbers} also includes the desiccation time $t_L$ (dashed lines).

\subsection{Bilobate shape}
Equation (\ref{eq:drdt}) can also be used with the full $r$ and $\theta$ dependent temperature field, although in its derivation lateral vapor flux was neglected.  Nevertheless, this still provides insight into how pronounced of a bilobate shape will emerge for the retained ice.

In non-dimensional variables $t' = t/t_D$ and $r'=r/R$, (\ref{eq:drdt}) becomes
\begin{equation}
\frac{dr'}{dt'} = \frac{\rho_s(T)}{6\rho_s(\bar T)} {1\over r'(1-r')}
\label{eq:drdtprime}
\end{equation}
where $r'$ and $T$ are both functions of $\theta$ and $t$. The solution $r'(\theta,t)$ only depends on $\bar T$, since $T(r',\theta)$ also follows from $\bar T$.

\begin{figure}
\includegraphics[width=7.5cm]{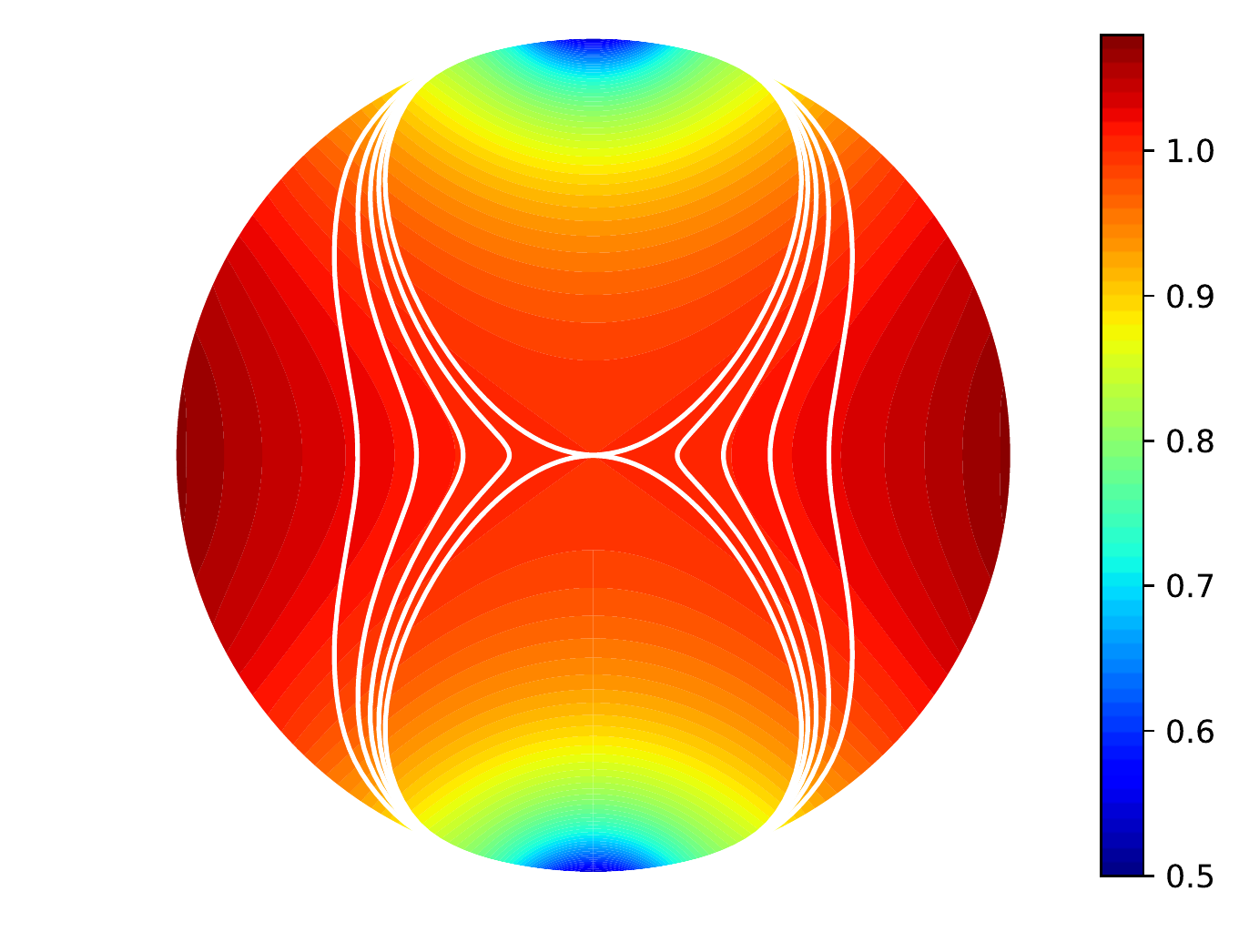}
\caption{Time and latitude dependent ice retreat according to equation (\ref{eq:drdtprime}) for $\bar T= 170$~K. White contours indicate ice table in time steps of $0.1t_D$, up to $0.5t_D$. Background color is temperature as in Figure~\ref{fig:Tsolution}b.}
\label{fig:dumbbell}
\end{figure}

Figure~\ref{fig:dumbbell} shows the numerically integrated results for the shape of the ice retained. The shallowest contour corresponds to $0.1 t_D$, and ice has retreated substantially at the equator, but not at the poles. Ultimately there is a pinch time $<t_D$ when the remaining ice is split into two hemispheric reservoirs of equal volume. Ice loss in the polar region is small even after a long time.  These cold spots can harbor near-surface ice reservoirs, as long as they have remained cold throughout the body's orbital history.

This pattern of desiccation is notable for MBCs, which may become active when collisions or mass shedding from rotational destabilization excavate near-surface ice \citep{hsieh09,hirabayashi2015_rotationalmassshedding,haghighipour2016_mbcimpacts}. It suggests that icy asteroids with sufficiently abundant ice to produce sublimation-driven activity close to their surfaces could be common, provided that the axis tilt is small and that activation events occur near the poles of such objects.

In the present model the regolith matrix is assumed to stay in place after the ice has retreated. If the ice fraction was higher, causing material to erode away as ice sublimated, potentially leaving behind an object with a dumbbell-like bilobate structure.  Several studies suggest that the formation of bilobate or contact binary structures observed for many comets and some small asteroids
\citep[e.g.,][]{harmon2010_8pcontactbinary,harmon2011_103pradar,magri2011_1996hw1contactbinary,jorda2016_67pshape,agarwal2017_binary288p}
are due to
the merger of a binary pair after tidal decay or
a low-speed collision of unrelated objects
\citep[e.g.,][]{magri2011_1996hw1contactbinary,taylor2014_tidalendstatebinaries,massironi2015_bilobate67porigin,jutzi2017_bilobate67porigin}. Differential sublimation rates could be an alternative mechanism for the formation of such structures.

As the ice retreats, the rotational inertia around the polar axis will become smaller than the rotational inertia of the other two principal axes, and the orientation of the object's rotation axis is expected to change.  On the other hand, small bodies are never perfectly spherical, and the shape defines a preferred rotation axis that is robust with respect to small changes in inertia. If the body does reorient, the polar region is quickly devolatilized.

\subsection{Application to specific populations}

Desiccation on a global scale for an entire body is governed by the temperature $\bar T$. It is not determined by the body's effective temperature, which is larger than $\bar T$, nor by perihelion temperatures, which are only relevant for relatively shallow depths. However, $\bar T$ depends on more than just an object's semi-major axis. The analysis presented in Fig.~\ref{fig:Tfroma}a shows how $\bar T$ also depends on thermal inertia and axis tilt.  A fast rotator with zero axis tilt will have the highest $\bar T$, and therefore serves as a ``hot'' end-member case.

\begin{figure}[tbh!]
a)\\
\includegraphics[width=7.5cm]{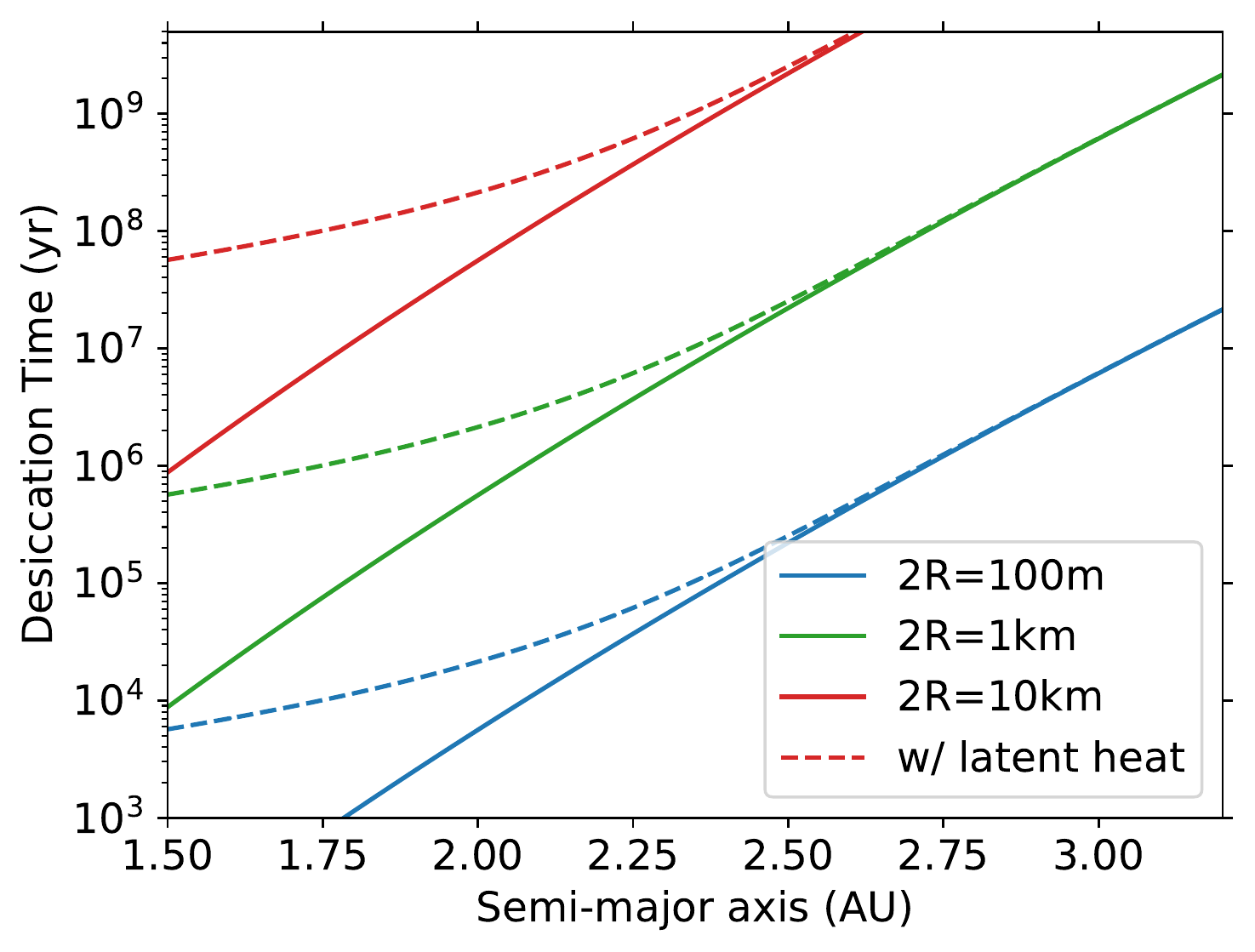}\\
b)\\
\includegraphics[width=7.5cm]{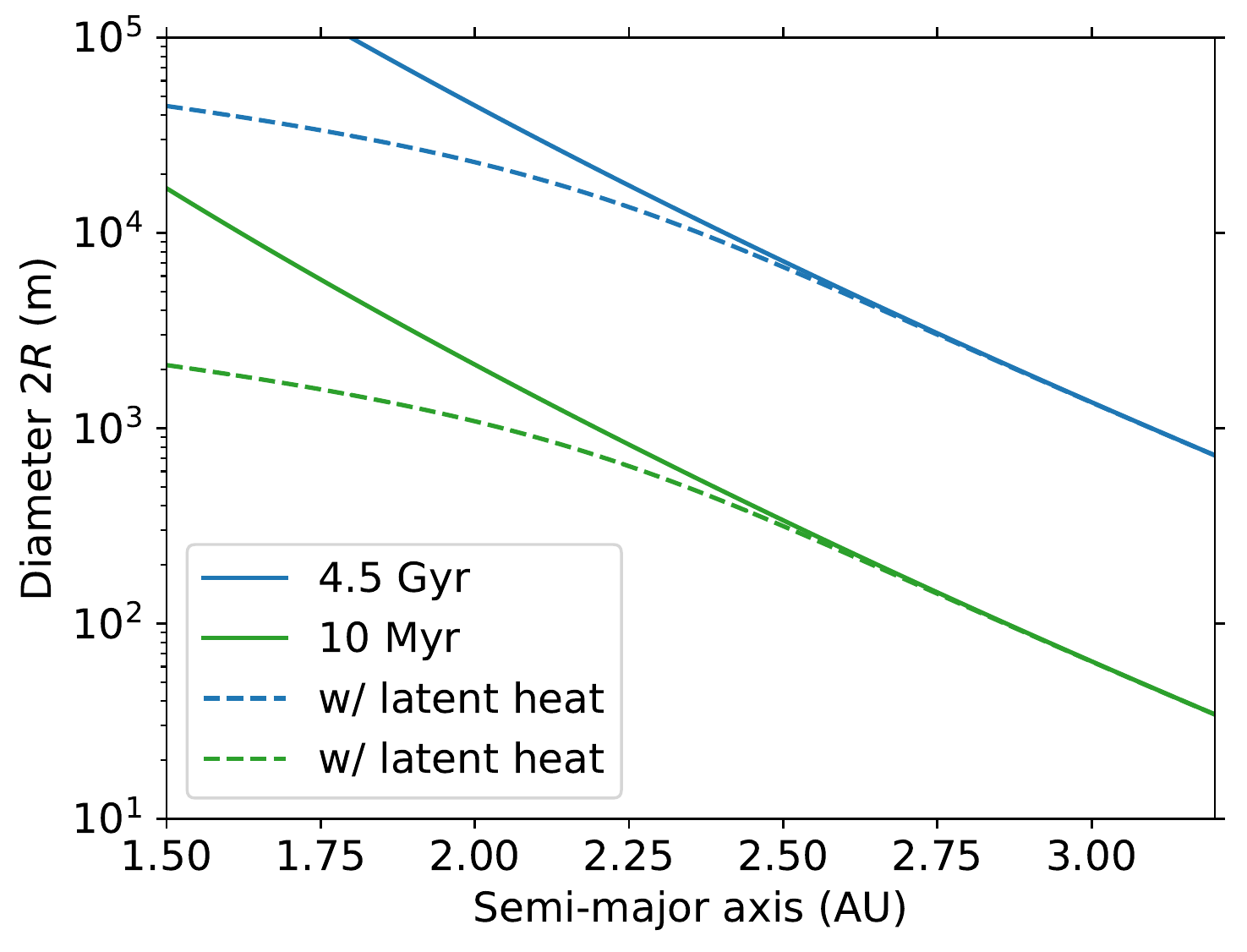}
\caption{Same as Figure~\ref{fig:thenumbers} but plotting desiccation time and threshold diameters for fixed desiccation times as functions of semi-major axis distance. See caption of Figure~\ref{fig:thenumbers} for further details. Plotted functions use $\bar T=T_u$, so represent pessimistically short desiccation times.
\label{fig:thenumbers2}}
\end{figure}

Figure~\ref{fig:thenumbers2} shows desiccation rates for this hot end-member case as a function of semi-major axis $a$.  As in Fig.~\ref{fig:thenumbers} these results assume a mean free path of $\zeta=1$~cm for water molecules and an albedo of 5\%. Desiccation times are proportional to $\zeta$ and $R^2$.

The Beagle family has an estimated age of 10~Ma and is centered at 3.157~AU with orbital eccentricities around 0.15, and contains the main-belt comet (MBC) 133P/Elst-Pizarro \citep{nesvorny08}. For $\zeta = 1$~cm, the size threshold for complete desiccation of a Beagle family asteroid is $R=41$~m. If we take $0.11 t_D$, when half the ice is lost, the threshold is 125~m. The smallest known Beagle family members are significantly larger than that \citep[$R\gtrsim1$~km, assuming albedos of $\sim$5\%;][]{nesvorny2015_hcmfamilies_pds}, so even for larger $\zeta$, all known Beagle family members should have retained most of their ice.

MBCs 133P, 176P, 238P, 288P, 313P, 324P, 358P, P/2013 R3, and P/2016 J1 have semi-major axes that range from 3.0~AU to 3.2~AU \citep{hsieh06,snodgrass17}.  Once the total insolation $\bar Q$ is adjusted for orbital eccentricity, the range of equivalent semi-major axes is still almost the same.  Objects with radii of $R\gtrsim 0.5$~km, which include all of the known MBCs, have desiccation time scales that exceed the age of the solar system. Many MBCs may be much younger, being apparent members of asteroid families formed in catastrophic disruption events of larger parent bodies in the relatively recent past \citep[cf.][]{hsieh2018_mbcfamilies}. Recent formation of MBCs is also supported by the fact that they exhibit comet-like activity, implying the presence of ice in near-surface layers and not only in their deep interiors.  Generally speaking, in the outer main belt ($a\gtrsim 2.8$~AU) bodies with radii larger than a few km ($R\gtrsim 3$~km for $\zeta=1$~cm) are expected to have retained ice.

Although collisionally sculpted, the current asteroid size distribution arose early in its history and there was little further evolution of the size distribution, which is nearly fossilized \citep{bottke05}.  Nevertheless, small bodies are continually formed through collisions and either gradually ejected or destroyed by further collisions.
\cite{granvik16} conducted Yarkovsky effect modeling on a larger number of asteroids, and found that most 100~m size objects were ejected over the integration period of 100~Myr. Hence, due to the Yarkovsky effect alone, asteroids of 100~m and smaller can be expected to be younger than 100~Myr. 
The typical lifetimes of 100~m asteroids against collisional destruction have been similarly estimated to be roughly 100~Myr \citep{bottke2005_astcollisionaldepletion}.
Objects of this size and age retained some of their ice at temperatures below about 160~K, which includes all bodies beyond a semi-major axis of 3.0~AU.  Hence, only in the outermost main belt are 100~m large asteroids expected to retain some ice, since they are likely to be $<$100~Myr old.

Objects with $a>2.5$~AU (i.e., in the middle and outer asteroid belt) and $R>7$~km should have been able to retain ice in their interiors over the age of the solar system, assuming that they formed with ice and that ice survived the period of radiogenic heating. This is valid for vapor mean free paths of up to $\zeta=1$~cm. (Ceres, at $a=2.77$~AU, retained ice within the top meter of the surface \citep{prettyman17} only because of its very low thermal inertia and small axis tilt \citep{schorghofer16a}.)

The NEO population covers a wide range of semi-major axes, but many NEOs originate in the main belt, and especially from the inner main belt ($a<2.5$~AU) \citep{binzel15rev}.
Nearly a thousand NEOs have diameters larger than 1~km, but most are smaller (https://cneos.jpl.nasa.gov/\allowbreak{}stats/\allowbreak{}size.html).
For an NEO with a diameter of 1~km or less to have retained ice in its interior, one of the following conditions would have to be met:
1) a semi-major axis in the outer belt or beyond,
2) a mantle of very low thermal inertia, which lowers interior temperature,
3) a young age due to a recent break-up from an ice-rich body, or
4) a stable and moderately small axis tilt that would maintain cold polar regions.

\section{Conclusions}

An idealized model of interior temperatures enables us to evaluate ice loss over a wide range of scales and parameters. For the surface temperature distribution of a fast rotator with a rotation axis perpendicular to the orbital plane, the interior equilibrium temperature is given by eqs.\ (\ref{eq:fullsolution}) and (\ref{eq:Crecursion}). Lateral temperature variations decrease with a depth-scale of about 3/10th of the body radius, and the average surface temperature $\bar T$ is representative of much of the body interior. This assessment also applies to thermally equilibrated spherical bodies with other surface temperature distributions, only $\bar T$ needs to be determined by other means than eqs.\ (\ref{eq:Teff},\ref{eq:TfromTeff}).

The surface-area-averaged and orbit-averaged surface temperature $\bar T$ plays a key role in assessing ice loss rates, because it represents the temperature of the body center. It depends, among other factors, on semi-major axis, thermal inertia, and spin axis tilt. Numerical exploration shows that a fast rotator with zero axis tilt represents a hot end member case, with a temperature about 1\% lower than the effective temperature, eq.\ (\ref{eq:TfromTeff}).

For a spherically averaged model, the time evolution of the ice retreat is obtained analytically (\ref{eq:tofr},\ref{eq:roft}), and the time until complete desiccation $t_D$ is given by eq.\ (\ref{eq:tD}); half of the ice mass is lost after $0.11 t_D$, and 7/8th are lost after $t_D/2$. Figure~\ref{fig:thenumbers} shows results for the desiccation time scale.

The model equations also enable us to investigate the latitude dependence of ice retreat. Figure~\ref{fig:dumbbell} illustrates the emergence of a bilobate structure. When the tilt of the rotation axis is not large, the polar regions retain subsurface ice long after the center of the body has lost its ice. Hence, cold polar areas may harbor ice on NEOs and small main belt asteroids that are otherwise devoid of ice.

The latent heat of the sublimating ice causes the ice to be colder than the surface average. Within a thermally equilibrated and spherically averaged model, this temperature difference is given by (\ref{eq:DeltaTfinal}) and is independent of body size and remains constant as the ice retreats toward the body center. These temperatures depend on the physical properties of the material, and for the values provided above, latent heat starts to significantly retard ice loss at mean surface temperatures above 190~K. Desiccation times for temperatures exceeding this regime are calculated based on numerical solutions to a nonlinear equation.

Applying these formulae we can make inferences about specific small body populations. First, all known Beagle family members should have been able to retain ice from their parent body over the age of the family ($\sim$10~Myr).  Next, in the outer belt, bodies with radii larger than a few km should be able to retain ice over the age of the solar system, and in the middle belt and beyond, bodies need to be nearly 10~km in diameter to have been able to retain most of their ice over this duration. These are conservative estimates because low thermal inertia and high axis tilt lower interior temperatures below the value used for these estimates and many dynamically younger objects are embedded in the main belt.
Lastly, NEOs (most of which are necessarily small) should only be able to retain ice under particularly favorable circumstances, such as very young ages, semi-major axes in the outer belt region, extremely low thermal inertia, or a small spin axis tilt and stable spin axis orientation.

\vspace{1em}
{\bf Acknowledgments:}
We thank Robert Jedicke, Karen Meech, and Abel M\'endez for insightful discussions.
This material is based upon work supported by the National Aeronautics and Space Administration under Grant No.\ 80NSSC17K0723 through the Solar Systems Working Program and through the NASA Solar System Exploration Research Virtual Institute 2016 (SSERVI16) Cooperative Agreement (NNH16ZDA001N).
No data products to report.


\end{document}